# Exploration of Speech Enabled Systems for English

**Kamlesh Sharma[1], T. Suryakanthi[2], Dr. T. V. Prasad[3]**
[1,2] *Research Scholar, Dept. of Comp. Sc. & Engg., Lingaya's University, Faridabad, India*
[3] *Dean of Computing Sciences, Visvodaya Technical Academy, Kavali, Andhra Pradesh, India*
[1]kamlesh0581@gmail.com
[2]ch.kanthi@gmail.com
[3]tvprasad2002@yahoo.com

*Abstract* — This paper presents exploration of speech enable operating systems, software, and applications. It begins with a description of how such systems work, and the level of accuracy that can be expected. It explains the applications of speech recognition technology in different areas education, medical, mobile computing, railway reservation, dictation, and web browsing. A brief comparison of the operating systems supported for voice, speech recognition software or tool. It gives the brief introduction about the potential of voice/speech recognition software. It explains the feature of different speech enable Operating system and speech recognition software. Windows speech recognition have many innovative features for Windows operating system and efficiently assist the computer to control, dictate, navigate, selecting the words, sending emails and correcting the words or sentences. It also explains the benefits and issue related to speech technology. In last era speech recognition technology grew tremendously. There are large number of companies who are working in these area and developing software for the people who are not able to control the system through keyboard or mouse such as physically impaired and senior citizens. This paper gives a brief introduction of speech enabled OS and speech recognition software.

## A. INTRODUCTION

Dictate to computer to send email, use applications, and have hands-free use of the computer. Talking into the computer is a futuristic technology available today (and rapidly improving in its accuracy), allowing to write with voice. Using voice recognition software can also help prevent or reduce RSI, Carpal Tunnel Syndrome. Speech recognition is a futuristic technology most suited to the fantastical suited for man machine relation. Table 3 lists 29 English voice/speech enable systems which shows that speech technology is largely viewing as an upcoming technology. Speech enable computer system entered in the each field in last 4 years education, medical, scientific and last year, the field has finally entered the mainstream, with major commercial applications in telephony, dictation, and PC command and control. Speech recognition is now viewed as a key enabling technology that provides a natural user interface to computer systems, and it is expected to revolutionize how people use their computers. Driven by rapidly dropping prices and improved accuracy rates, the market for speech-recognition products is projected to grow in million last years. Major hardware and software manufacturers are playing a big role. The speech recognition software is "the next big wave" in computer industry. The intends to include speech-recognition systems among the products in the near future, beginning with the operating system used in palm-top computers are another significant move. [1]

Moreover, sophisticated operating systems increase the efficiency and consequently decrease the cost of using a computer [3]. A large number of operating systems of various types are available for both research and commercial purposes, and these operating systems vary greatly in their structures and functionalities.

Computers have progressed and developed so have the operating systems. There is list of the different operating systems which comes in the category of speech enabled systems and large number of examples of speech recognition systems. With the integration of computers and telecommunications, the mode of information access becomes an important issue. The designs of the prevalent human machine interfaces are more suitable for easier interpretation of information by computers than by human beings. The concept of machine being able to interact with people in a mode that is natural as well as convenient for human beings.

## B. SPEECH/VOICE ENABLED OPERATING SYSTEMS

Speech recognition is the latest feature of the operating system. This allows the user to both enter commands, such as "open file", and to dictate text straight into an application. Added speech functionality enables the translation of text between a small numbers of languages. The table 2 in appendix compares different feature of Operating System.

1. *Mac OS X Lion*

Speech recognition in Mac OS X Lion enables the Mac to recognize and respond to human speech. The only thing need to use is a microphone, and all laptops and iMacs have a built-in mic these days, Speech Recognition lets it issue verbal commands such as "Get my mail!" to your Mac and have it actually get your e-mail. It can also create AppleScripts and then trigger them by voice. [4]

2. *Windows Vista*

Windows Speech Recognition is a new feature in Windows Vista, built using the latest Microsoft speech technologies. Windows Vista Speech Recognition provides excellent recognition accuracy that improves with each use as it adapts to the speaking style and vocabulary. Speech Recognition is available in English (U.S.), English (U.K.), German (Germany), French (France), Spanish (Spain), Japanese, Chinese (Traditional), and Chinese (Simplified).

3. *Windows XP*

Windows Speech Recognition is a new feature in Windows Vista, built using the latest Microsoft speech technologies. Windows Vista Speech Recognition provides excellent recognition accuracy that improves with each use as it adapts to the speaking style and vocabulary. Speech Recognition is available in English (U.S.), English (U.K.), German (Germany), French (France), Spanish (Spain), Japanese, Chinese (Traditional), and Chinese (Simplified). [5]

4. *Windows 7*

The Windows Speech Recognition by Microsoft is the speech recognition system that comes built intoWindows 7. Windows Vista and Windows 7 include version 8.0 of the Microsoft speech recognition engine. Speech Recognition is available only in English, French, Spanish, German, Japanese, Simplified Chinese, and Traditional Chinese. Windows 7 that enables users to control the mouse cursor and keyboard through speech recognition.[1] Voice Finger improves on the default Windows Speech Recognition tools by reducing the number or length of voice commands required to carry out various tasks.

C. VOICE/SPEECH RECOGNITION SYSTEMS (SOFTWARE)

There are large numbers of speech recognition systems of different kinds that are commercially available. The Table 3 and 4 lists details of 29 systems and 5 speech tools. A Voice–user interface (VUI) makes human interaction with computers possible through a voice/speech platform in order to initiate an automated service or process.

A VUI is the interface to any speech application which controlled a machine by simply talking to it. Until recently, this area was considered to be artificial intelligence. However, with advances in technology, VUIs have become more commonplace, and people are taking advantage of the value that these hands-free, eyes-free interfaces provide in many situations.VUIs need to respond to input reliably, or they will be rejected and often ridiculed by their users. Designing a good VUI requires interdisciplinary talents of computer science, linguistics and human factors psychology – all of which are skills that are expensive and hard to come by. Even with advanced development tools, constructing an effective VUI requires an in-depth understanding of both the tasks to be performed, as well as the target audience that will use the final system. The closer the VUI matches the user's mental model of the task, the easier it will be to use with little or no training, resulting in both higher efficiency and higher user satisfaction. [6]

The characteristics of the target audience are very important. For example, a VUI designed for the general public should emphasize ease of use and provide a lot of help and guidance for first-time callers. In contrast, a VUI designed for a small group of power users (including field service workers), should focus more on productivity and less on help and guidance. Such applications should streamline the call flows, minimize prompts, eliminate unnecessary iterations and allow elaborate "mixed initiative dialogs", which enable callers to enter several pieces of information in a single utterance and in any order or combination. In short, speech applications have to be carefully crafted for the specific business process that is being automated. [7]

Not all business processes render themselves equally well for speech automation. In general, the more complex the inquiries and transactions are, the more challenging they will be to automate, and the more likely they will be to fail with the general public. In some scenarios, automation is simply not applicable, so live agent assistance is the only option. A legal advice hotline, for example, would be very difficult to automate. On the flip side, speech is perfect for handling quick and routine transactions, like changing the status of a work order, completing a time or expense entry, or transferring funds between accounts.

Speech recognition (also known as automatic speech recognition, computer speech recognition, speech to text, or just STT) converts spoken words to text. The term "voice recognition" is sometimes used to refer to recognition systems that must be trained to a particular speaker—as is the case for most desktop recognition software. Recognizing the speaker can simplify the task of translating speech.

Speech recognition is a broader solution that refers to technology that can recognize speech without being targeted at single speaker—such as a call system that can recognize arbitrary voices.

Speech recognition applications include voice user interfaces such as voice dialing (e.g., "Call home"), call routing (e.g., "I would like to make a collect call"), domestic appliance control (Washing Machine, Mixer Grinder), search (e.g., find a podcast where particular words were spoken), simple data entry (e.g., entering a credit card number), preparation of structured documents (e.g., a radiology report), speech-to-text processing (e.g., word processors or emails), and aircraft (usually termed Direct Voice Input).

Voice interfaces have their greatest potential in the following cases, which make relying on the traditional keyboard-mouse-monitor combination problematic:

- Users with various disabilities, who cannot use a mouse and/or a keyboard or who cannot see pictures on the screen. Voice output is the main way for visually impaired users to interact with computers, and because these users rely so heavily on audio presentation of information, it is very important to design Web pages with voice-only browsers in mind.

- Users who are in an eyes-busy, hands-busy situation. Whether or not they have disabilities, the keyboard-mouse-monitor combo fails users in these situations, such as when they're driving cars or repairing complex equipment.

- Users who don't have access to a keyboard and/or a monitor. In this case, users might, for example, access a system by payphone.

- User who are not aware of computer skills. In this case can access the system by voice commands and make the system to execute them.

### D. FEATURES OF SPEECH RECOGNITION SOFTWARE

Speech/voice enabled systems provides a set of features that efficiently assists user in controlling the computer by voice. Whether users are starting an application, selecting a word, or correcting a sentence, they are always in control and speech recognition features guides user to completing the task. Table 1 shows different key feature of speech/voice enable technology.

TABLE 1
KEY FEATURES OF SPEECH ENABLED SYSTEM

| S. No. | Feature | Description |
|---|---|---|
| 1 | Commanding | Commands allow controlling the system and applications, such as opening, closing, sending emails, reading emails and browse the internet. |
| 2 | Dictating | Dictate emails and documents and make corrections and save the work by voice. |
| 3 | Calculating | Calculate mathematical expressions, such as addition, subtraction, multiplication and division. |
| 4 | Sending/Receiving Mail. | Sending and receiving mails through the voice. |
| 5 | Fax-sending | Sending Fax and receiving Fax. |
| 6 | Correcting | Correct the words which are incorrectly recognized by selecting form alternatives for the dictated phrase or word or by spelling the word. |

### E. BENEFITS FROM USING SPEECH RECOGNITION

The system will be become a lot more productive since they will be able to get the work done more quickly as most of them can dictate at least 3 times faster than users can type. It will be like having a Personal Assistant available 24/7. Benefit from immediate transcription results - no more waiting for the work to be typed. If user suffers from RSI or arthritis, there will be much less strain on the hands

Anyone who relies heavily on a PC or Mac will be a lot more productive including Teachers, Lecturers, Students Special Educational Needs Students, Directors, Managers, Office staff, Administrators, Mobile workers, Solicitors, Barristers, Lawyers, Legal Secretaries, Court Clerks, Surveyors, Building Engineers, Lecturers, Teachers and Students, Accountants and Financial Advisers, Administrators and users with disabilities such as repetitive strain injury (RSI), dyslexia, vision impairment and others

### F. ISSUES OF SPEECH RECOGNITION

There are many issues with voice-recognition/speech recognition software.

- Body noise: when user speaks in front of the speech recognition system, it receives the body noise such as hand movement, leg movement, laughing, sneezing, cough etc. which creates a problem for recognition of the correct word or phrase.

- Background noise: Background noise of the environment where the system is working, such as noise of fan, movement of chairs, gossiping of persons, movement of doors creates a problem at the time of recognition.

- Low quality microphone: A poor quality microphone without noise cancelling feature creates problem in recognition.

- Different languages: There are several languages across the world and so much variation in the syntax and vocabulary. People in different regions speak the same language but with different accents. People speak at different speeds, tones and pitches. Colloquialisms are used where the phrases and words can have several meanings. So recognizing the exact word by the speech system is difficult at times.

- Inconsistence speaking: People have their own voice modulation like speaking too loudly or too softly, which makes it difficult for the system to recognize what was said. Speaking at a consistent rate, without speeding up and slowing can reduce the problem.

- Speak without pause: This is a problem in which user speak words without pause which makes it difficult for the system to recognize the words.

- Pronunciation: This is a problem in which pronunciation of the words are not clear and makes it difficult for system to recognize the correct word. For example, sounding out each syllable in "pro-nun-ci-ation," will make it harder for the computer to recognize what user said. [2]

## VI. CONCLUSION

In this paper we have presented the exploration of speech enabled operating Systems and speech enabled system. We describe the features of speech enabled systems and issues or benefits of speech recognition systems. It is a paper for awareness of voice/speech based system. Given the current state of IT market and the research field, speech/voice based system will be the upcoming field of system. They are used to provide a bridge between the physically disable people and the voice/speech enable system. This kind of software's been developed to provide a fast method of writing on computer and can help people with a variety of disabilities. It is useful for people with physical disabilities who often find typing difficult, painful or impossible. Voice-recognition software can also help those with spelling difficulties, including users with dyslexia, because recognised words are almost always correctly spelled.

# Appendix A

TABLE 2
COMPARISON OF DIFFERENT FEATURES OF OPERATING SYSTEM

| System | Connectivity | Stability | Scalability | Multiuser | Multiplatform | Speech Enable | Non-Proprietary |
|---|---|---|---|---|---|---|---|
| **Legacy System** | Poor | Good | Medium-Huge | Yes | No | No | No |
| **MS-DOS** | None | Poor | Small | No | No | No | No |
| **Windows 3.x** | Poor | Poor | Small | No | No | No | No |
| **Windows95** | SMB Only | Fair | Small | Insecure | No | No | No |
| **WindowsNT** | SMB+ | Fair | Small-Medium | Yes | Yes, 2 | No | No |
| **WindowXP** | Excellent | Excellent | Small-Huge | Yes | Yes, Many | Yes | No |
| **UNIX** | Excellent | Excellent | Small-Huge | Yes | Yes, Many | No | No |
| **Linux** | Excellent | Excellent | Small-Huge | Yes | Yes, Many | No | No |
| **Mac OS** | Excellent | Excellent | Small-Huge | Yes | Yes | Yes | No |
| **Windows 7** | Excellent | Excellent | Small-Huge | Yes | Yes, Many | Yes | No |
| **Window 8** | Excellent | Excellent | Small-Huge | Yes | Yes, Many | Yes | No |

TABLE 3
DESCRIPTION OF ENGLISH VOICE/SPEECH RECOGNITION SOFTWARE

| S. No. | Name | Description | Developed by | Year |
|---|---|---|---|---|
| 1 | CMU Spimhinx | CMU Sphinx is speech recognition system that includes a series of Sphinx 2-4 and acoustic modael trainer(SphinxTrain).Speech decoder for speech recognition come with sample application and acoustic model.These have basically three major components acoustic model, language model and public domain pronunciation dictionary. | Carnegie Mellon University | 2000 |
| 2 | Julius | Julius is a high-performance, two-pass large vocabulary continuous speech recognition (LVCSR) decoder software for speech which can perform almost real-time decoding on most current PCs in 60k word dictation task using word 3-gram and context-dependent HMM. The main platform is Linux and other Unix workstations, and also works on Windows. Julius is open source and distributed with a revised BSD style license. | Japan | 1997 |
| 3 | Simon | The project provides a ready-to-use interface for the julius CSR engine for a handicapped child which is not able to use the keyboard well. It integrates into X11 and Windows and designed to be very flexible and allows customization for any application where speech recognition is needed. It can control a number of programs like a web browser , e-mail client, media center etc. It is possible to listen to music, watch a slide show, TV or videos or listen to the radio just with a few - free eligible - words like "right", "left", "up", "down", "ok", "stop" etc. | Simon Listens | 2007 |
| 4 | Iatros | iATROS is a system which is use for speech and handwriting input. The iATROS system is developed in a modular manner, with a core recognition engine and several utility functions that can be used in the construction of speech and handwriting-based applications, including multimodal and interactive applications.[8] | Universidad Politécnica de Valencia | 2006 |
| 5 | RWTH-ASR-QPL | RWTH ASR (RASR) is an open source speech recognition toolkit. RWTH ASR includes tools for the development of acoustic models and decoders as well as components for speaker adaptation, speaker adaptive training, unsupervised training, discriminative training, and word lattice processing. The software runs on Linux and Mac OS X. | RWTH Aachen University | 2009 |
| 6 | VoxForge | VoxForge is a free speech corpus and acoustic model repository for open source speech recognition engines. VoxForge was set up to collect transcribed speech to create a free GPL speech corpus for use with open source speech recognition engines. The speech audio files will be 'compiled' into acoustic models for use with open source speech recognition engines such as Julius, ISIP, and Sphinx and HTK . VoxForge use LibriVox as a source of audio data. | GNU General Public License | 1998 |

| S. No. | Name | Description | Developed by | Year |
|---|---|---|---|---|
| 7 | Dragon Dictate | Dragon Dictation is a speech recognition App for Apple's iOS platforms including iPhone, iPod touch and iPad. The App provides automatic speech-to-text capabilities. It works as an online solution, requiring an Internet connection by the user. | Nuance Communications | 2009 |
| 8 | iListen | iListen, is a speech recognition program for the Apple Macintosh. iListen is currently the only third-party software that allows inputting text using one's voice that works on newer Macintosh models | Mech Speech Inc. | 2006 |
| 9 | MacSpeech Dictate | MacSpeech Dictate was a speech recognition program. It used the Dragon speech recognition engine. MacSpeech Dictate Medical, a version with specialized vocabularies for doctors and dentists, was released in June 2009. MacSpeech Dictate Legal, with specialized vocabulary for lawyers, was released in July 2009. MacSpeech Dictate International, with support for speech recognition in English, French, German and Italian, was released in September 2009. Localized versions of MacSpeech Dictate are available in German, French and Italian. | Mech Speech Inc. | 2008 |
| 10 | MacSpeech Scribe | MacSpeech Scribe is speech recognition software for Mac OS X designed specifically for transcription of recorded voice dictation. It runs on Mac OS X 10.6 Snow Leopard. The software transcribes dictation recorded by an individual speaker. Typically the speaker will record their dictation using a digital recording device such as a handheld digital recorder, mobile smartphone (e.g. iPhone), or desktop or laptop computer with a suitable microphone. MacSpeech Scribe supports specific audio file formats for recorded dictation: .aif, .aiff, .wav, .mp4, .m4a, and .m4v. | Mech Speech Inc | 2009 |
| 11 | Speakable item | Speakable items is part of the speech recognition feature in the Mac OS and Mac OS X operating systems. It allows the user to control their computer with natural speech, without having to train the computer beforehand. The commands must be present in the Speakable items folder though but can be created with something as simple as a shortcut, AppleScript, keyboard command, or Automator workflows. | Apple Inc. | 1993 |
| 12 | ViaVoice | ViaVoice is language-specific continuous speech recognition software designed primarily for use in embedded devices. It can scan text directly from words, recognize, correct the word and popularly used for typing. | IBM | 2005 |
| 13 | Dragon Dictation | DragonDictate and Dragon Dictate are proprietary speech recognition software. DragonDictate for Windows was the original speech recognition application from Dragon Systems and used discrete speech where the user must pause between speaking each word. It can able to scan text directly from words, recognize, correct the word and popularly used for typing. | Nuance | |
| 14 | Siri Personal Assitant | Siri is an intelligent personal assistant and knowledge navigator which works as an application for Apple's iOS. The application uses a natural language user interface to answer questions, make recommendations, and perform actions by delegating requests to a set of Web services. It can perform tasks such as finding recommendations for nearby restaurants, or getting directions. | Siri, Inc | 2010 |
| 15 | VoiceAttack | VoiceAttack will take commands that you speak into your microphone and turn them into a series of keyboard key presses (and do other things like launch programs). VoiceAttack is designed to make games and applications more funny to use by adding voice as an extra controller.[9] | | |
| 16 | Voice Finger | Voice Finger is a software tool for Windows Vista and Windows 7 that enables users to control the mouse cursor and keyboard through speech recognition.[1] Voice Finger improves on the default Windows Speech Recognition tools by reducing the number or length of voice commands required to carry out various tasks.[10] | Robson Cozendey | 2009 |
| 17 | Trigamtech | Trigamtech is a speech recognition software which is used for medcal dictation and document management for health care.[11] | Trigram Technology | 1996 |
| 18 | Vocola | Vocola 3is a tool which works with Windows Speech Recognition, a free component of Windows Vista and Windows 7. WSR is a good recognition engine, with comparable speed and accuracy to Dragon NaturallySpeaking (DNS). While WSR lacks some of the polish and documentation of DNS, it is eminently usable. Vocola addresses some important limitations of WSR; in particular it supports dictation to any application. | Rick Mohr | 2002 |

| S. No. | Name | Description | Developed by | Year |
|---|---|---|---|---|
| 19 | Dragon Naturally Speaking | Dragon NaturallySpeaking is a speech recognition software package. dictation, text-to-speech and command input. The user is able to dictate and have speech transcribed as written text, have a document synthesized as an audio stream, or issue commands that are recognized as such by the program. In addition, voice profiles can be accessed through different computers in a networked environment, although the audio hardware and configuration must be identical on both machines. | Nuance Communication | 2011 |
| 20 | SpeechMagic | SpeechMagic features large-vocabulary speech recognition which is basically used for capturing information in a digital format. SpeechMagic supports 25 recognition languages and provides more than 150 *ConTexts* (industry-specific vocabularies). The technology is mainly used in the healthcare sector, however, applications are also available for the legal market as well as for tax consultants. | Nuance Communication | 2008 |
| 21 | VoxCommando | VoxCommando is a speech recognition and command utility that is developed for media PC. Control XBMC, iTunes, MediaMonkey and much more with speech. VoxCommando is different from other speech recognition applications in that it is extremely customizable.[13] | VoxCommando | |
| 22 | Tazti | is a speech recognition software package which supports Windows 7[2] 64-bit and 32-bit editions as well as Windows Vista and Windows XP. Dictation, voice search, PC video game play by voice and user configurable speech commands are among Tazti's many features. | Voice Tech Group, Inc. | 2011 |
| 23 | e-Speaking | A comprehensive voice and speech recognition program to use your voice for command & control of your computer and dictation. Reduce or eliminate mouse clicks or keyboard input. Open Web sites, documents, or programs using your voice. Perform navigation and editing functions, dictate letters, memos, and email messages. Begin talking to your PC now. With Animated Avatar. Hear and see your computer read emails, documents, and memos. | e-speacking | 2009 |
| 24 | Clap Commander | Clap Commander is an easy-to-use program that allows you to control your computer from a distance by clapping your hands. Clap once and Media Player will launch, clap twice and make Media Player pause or start a movie over from the beginning, clap three times and the computer will turn off. The program recognizes the sounds picked up by the microphone and if they resemble a person clapping, a user-defined action will be performed | MentalWays.com | 2009 |
| 25 | Loquendo ASR | Loquendo ASR is speech recognition software based on advanced neural network technologies, along with excellent robustness to background noise, it guarantees high performance even on large-scale vocabulary speech, whether using speech grammars or statistical language models. All this enables a dialogue with the user which is simple and natural. ideal for any application environment, whether Telephony, Web, Desktop, Embedded, Automotive or Mobile, thanks to specific product profiles, and is equipped with specialized acoustic models to always guarantee maximum performance whatever the application context. Furthermore, because it is speaker independent, it is ready-to-use, requiring no prior training. | Naunace Inc. | 2010 |
| 26 | Simmortel Voice | Simmortel Voice Technologies is a computer software technology startup,[1] started from IIT Kanpur,[2] India, that provides hosted telephony, voice and automatic speech recognition solutions. It provides a platform for hosting Interactive Voice Response Systems, speech recognition,[3] and call center applications.[15] | IIT Kanpur | 2007 |
| 27 | Tellme Networks | "Tell Me", provided time-of-day announcements, weather forecasts, brief news and sports summaries, business searches, stock market quotations, driving directions, and similar amenities. Operating by voice prompts and speech-recognition software. The 800-555-Tell (800-555-8355) information number, identifying itself as "Tell Me", | subsidiary of Microsoft | 1999 |
| 28 | Vocapia Research | Vocapia Research develops technologies for multilingual, large vocabulary speech recognition (also called speech-to-text conversion), automatic audio segmentation, language identification and speaker recognition. Vocapia's VoxSigma™ speech-to-text software suite delivers state-of-the-art performance in many languages for a variety of audio data types, including broadcast data, parliamentary hearings and conversational speech. Vocapia | Vocapia Research | 2000 |

| S. No. | Name | Description | Developed by | Year |
|---|---|---|---|---|
| | | Research, formerly *Vecsys Research*[1], is a high tech research and development company (R&D), developing technologies for multilingual, unconstrained speech-to-text transcription systems, automatic segmentation of audio data, as well as language and speaker recognition. | | |
| 29 | Google Voice Search | Google Voice Search or Search by Voice is a Google product that allows someone to use Google Search by speaking on a mobile phone or computer, i.e. have the device search for data upon entering information on what to search into the device by speaking.There is also Voice Action which allows one to give speech commands to an Android phone. This feature is however currently limited to only a few languages: U.S. and British English, French, Italian, German and Spanish. | Google | 2011 |

TABLE 4
DESCRIPTION OF VOICE/SPEECH RECOGNITION TOOLS

| S. No. | Name | Description | Developed by | Year |
|---|---|---|---|---|
| 1 | AT&T Watson API | AT&T WatsonSM speech recognition technology APIs for developers to incorporate the speech engine's capabilities into their own products with minimum hassle. The AT&T Watson speech recognition technology APIs brings speech to web search, local business search, question and answer, voicemail to text, SMS, U-verse electronic programming guide and other generic speech functions.[14] | AT& T Lab | 2012 |
| 2 | CSLU Toolkit | The CSLU Toolkit is a software library comprising a comprehensive suite of tools that enable exploration, learning, and research into speech and human-computer interaction. The CSLU Toolkit have the feature of speech recognition, natural language understanding, speech synthesis and facial animation technologies. The toolkit provides a comprehensive, powerful and flexible environment for building interactive language systems. The tools include Audio, Display, Speech recognition, Speech generation, Animated faces.[16] | Technology and Research Collaborations, Oregon Health & Science University | 2004 |
| 3 | HTK | HTK (Hidden Markov Model Toolkit) is software toolkit for handling HMMs. It is mainly intended for speech recognition, but has been used in many other pattern recognition applications that employ HMMs, including speech synthesis, character recognition and DNA sequencing.The Hidden Markov Model Toolkit (HTK) is a portable toolkit for building and manipulating hidden Markov models. HTK is primarily used for speech recognition research although it has been used for numerous other applications including research into speech synthesis, character recognition and DNA sequencing.[12] | Machine Intelligence Laboratory | 1993 |
| 4 | iSpeech ASR API | The iSpeech API allows developers to implement Text-To-Speech (TTS) and Automated Voice Recognition (ASR) in any Internet-enabled application.The API's are platform agnostic which means any device that can record or play audio connected to the Internet can use the iSpeech API. | iSpeech Inc. | 2011 |
| 5 | Naunace Recognizer | Nuance Recognizer is the software at the core of our contact center automation solutions. Built upon years of experience across six different product lines, our tenth-generation Automatic Speech Recognition (ASR) engine is used around the world in over 79 different languages. It delivers the industry's highest recognition accuracy even as it encourages natural, human-like conversations that create more satisfying self-service interactions with your customers. | Naunace Inc. | 2008 |